# Magnetic phase transitions and superconductivity in strained FeTe

A. Ciechan,[1] M. J. Winiarski,[2] and M. Samsel-Czekała[2]

[1]*Institute of Physics, Polish Academy of Sciences, al. Lotników 32/46, 02-668 Warsaw, Poland*
[2]*Institute of Low Temperature and Structure Research,*
*Polish Academy of Sciences, ul. Okólna 2, 50-422 Wrocław, Poland*

The influence of hydrostatic pressure and *ab*-plane strain on a magnetic structure of FeTe is investigated from first principles. The results of calculations reveal a phase transition from the antiferromagnetic double-stripe ordering at ambient pressure to the ferromagnetic ordering at 2 GPa or under compressive strain reducing the lattice parameter *a* of about 3%. In turn, the tensile strain less than 2% induces the phase transition to the antiferromagnetic single-stripe ordering. It corresponds to the superconducting FeTe thin films, thereby confirming that the superconducting state is positively linked to the single-stripe antiferromagnetic fluctuations. Both types of transitions indicate that the position of Te atoms in the crystal is crucial for magnetic and superconducting properties of iron chalcogenides.

## I. INTRODUCTION

Fe(Se,Te) systems exhibit the complex magnetic and superconducting phase diagrams. FeSe is a superconductor with critical temperature $T_c$ =8 K [1], which raises upon both a partial tellurium substitution and hydrostatic pressure. The FeSe$_{1-x}$Te$_x$ solid solutions are superconducting for $x$ <0.8 and maximum $T_c$ =15 K is observed for $x$ =0.5 [2–4]. In turn, the end member FeTe ($x$ =1) displays the antiferromagnetic (AFM) ground state. Interestingly, there are theoretical predictions that FeTe could be a superconductor with even higher $T_c$ than FeSe [5, 6]. It is likely because of the similar electronic band structures of both compounds and also the same Fermi surface nesting, which might induce spin fluctuations and enhance superconductivity. However, such fluctuations correspond to a single-stripe AFM order with the $(\pi, \pi)$ wave vector, while FeTe is a double-stripe AFM material with the $(\pi, 0)$ propagation vector, turning into the incommensurate $(\delta\pi, 0)$ vector in the iron-rich samples [7–10].

For superconducting Fe(Se,Te) the critical temperature values raise with external pressure up to 37 K [11–14]. Although the resistivity of the pure FeTe at room temperature decreases with increasing pressure [15], the superconducting phase for this telluride does not appear even under as high pressure as 19 GPa [16]. Instead, two high-pressure magnetic phases in FeTe were observed, which is probably linked to an enhancement of the Te-Te hybridization between neighbouring Fe-Te layers. Recently, the transition from a low pressure AFM phase to a high pressure ferromagnetic (FM) phase in Fe$_{1.03}$Te has been reported [17].

Values of $T_c$ of the iron chalcogenide superconductors can be tuned by non-hydrostatic pressure in lattice mismatched epitaxial films. The tensile strain suppresses superconductivity of FeSe on MgO and SrTiO$_3$ [18], while the compressive biaxial (*ab*-plane) or uniaxial (*c*-axis) strain on Fe(Se,Te) causes the increase of the $T_c$'s [19–21]. The method is successful also in the pure FeTe, where superconductivity emerges in thin films as an effect of tensile stress [22].

Both magnetic and electronic transport properties of Fe(Se,Te) systems are strongly related to their crystal structure. At room temperature, the compounds are tetragonal [2]. For superconducting FeSe, the orthorhombic distortion occurs upon cooling to low temperature of about 70-90 K [2, 23]. In FeTe, the monoclinic lattice distortion takes place below the Neel temperature of 70 K [2, 7, 8, 24]. The results of crystal structure refinements of the alloys indicate also that some of the distances and angles are more preferable in the superconducting state [12, 14, 22, 25, 26].

In this work, the magnetic phase transitions of FeTe are obtained by *ab initio* calculations under various stress conditions. The AFM to FM switch emerges at pressure of 2 GPa, being in excellent agreement with the experimental value [17], as well as under *ab*-plane compressive stress reducing the lattice parameter *a* by about 3%. In turn, the *ab*-tensile strain of less than 2% changes the AFM order from the double- to single-stripe one. Since negative biaxial pressure induces superconductivity in FeTe thin films grown on SrTiO$_3$ or MgO [22], it confirms that superconducting state is positively linked to the single-stripe AFM arrangement, but not to the double-stripe AFM or FM arrangements. A further analysis shows that both transitions are related to the variation of the chalcogen atomic position, which strongly modifies the electronic and magnetic properties of the system.

## II. COMPUTATIONAL METHODS

The examination of the pressure induced magnetic transition of FeTe was performed within the density functional theory (DFT) [27] in the generalized-gradient approximation (GGA) of the exchange-correlation potential [28]. We used the pseudopotential method, based on plane-waves and Projector-Augmented Waves, imple-mented in the QUANTUM ESPRESSO code [29]. The kinetic energy cutoff for wavefunctions and for charge density were 35 Ry and 210 Ry, respectively. The calculations were performed with $12 \times 12 \times 8$ k-point grid in the non-equivalent part of the Brillouin zone ($6 \times 12 \times 8$ grid for $2a \times a \times c$ supercell and $8 \times 8 \times 8$ grid for $\sqrt{2}a \times \sqrt{2}a \times c$ supercell). The optimization of the lattice structures was carried out with $10^{-3}$ Ry/Bohr convergence criterion on forces.

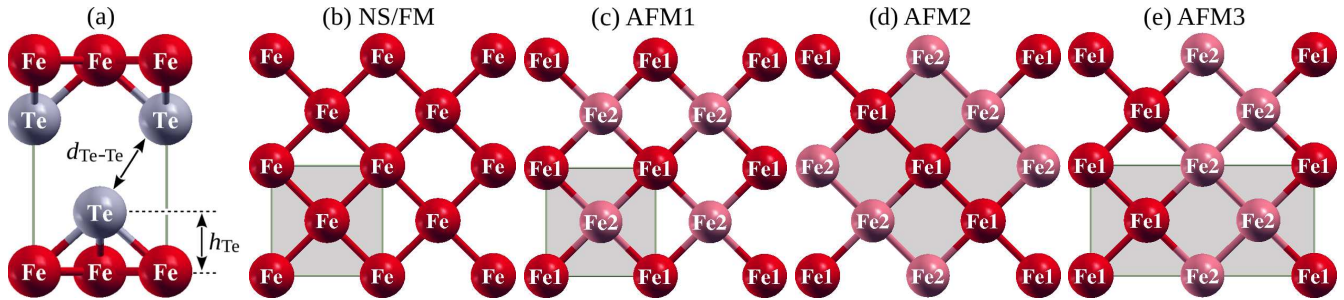

FIG. 1: (a) A side view of FeTe tetragonal structure. The $h_{Te} = z_{Te}c$ denotes the Te-distance from the Fe plane, $d_{Te-Te}$ - the interlayer Te-Te distance. Schematic arrangements of in-plane Fe spins displaying possible magnetic orders in FeTe: (b) non-magnetic (NS) or ferromagnetic (FM) and some antiferromagnetic ones: (c) checkerboard (AFM1), (d) single stripes (AFM2) and (e) double stripes (AFM3). The Fe atoms with different spin directions are indicated as Fe1 and Fe2. The magnetic unit cell in each case is marked by the shaded area.

We start with the tetragonal phase of the PbO-type (space group $P4/nmm$) and experimental lattice parameters [15]. The Fe ions form a square lattice and the Te ions are located either directly above or below the Fe plane (figure 1a). Thus, the unit cell contains 4 atoms in positions: Fe$(0,0,0)$, Fe$(0.5,0.5,0)$, Te$(0.5,0,z_{Te})$ and Te$(0,0.5,-z_{Te})$. Since FeTe displays spin ordering changes under pressure [16, 17], we take into account FM and some possible AFM spin arrangements suggested for chalcogenides in literature [30–32]. All considered here magnetic structures in the $ab$-plane are shown in figures 1(b-e). In the single unit cell, FM and the checkerboard configuration (denoted as AFM1) can be realized. A single-stripe spin arrangement (AFM2), which requires the $\sqrt{2}a \times \sqrt{2}a \times c$ unit cell, was previously predicted as a ground state of FeSe. The last one, i.e. the double stripe AFM configuration (AFM3), realized in the $2a \times a \times c$ unit cell, was found out for FeTe at ambient pressure both theoretically and experimentally [7, 8, 30, 31].

Next, we test the magnetic phases of FeTe under pressure and $ab$-plane strain. Since not all experimental structure parameters of the considered system were available in the literature, both lattice parameters and atomic positions have been optimized. Under hydrostatic pressure some magnetic cells distort from initial tetragonal to orthorhombic or monoclinic ones. For biaxial strain, we fixed the lattice parameter $a$ and tetragonal cell of the PbO-type to simulate epitaxial films.

We discuss electronic band properties of such optimized FeTe under various conditions and check if the similar phase diagrams are obtained as an effect of vary-ing only the $z_{Te}$ position in the crystal.

TABLE I: The total energies per formula unit of the spin-polarized states with respect to the non-spin-polarized one ($\Delta E$) and the magnetic moment of the Fe atoms ($\mu$) obtained for the considered spin configurations of FeTe. The experimental tetragonal crystal structure is assumed [15].

| In-plane | parallel spins along $c$ | | antiparallel spins along $c$ | |
| --- | --- | --- | --- | --- |
| | $\Delta E$ (meV/f.u.) | $\mu$ ($\mu_B$) | $\Delta E$ (meV/f.u.) | $\mu$ ($\mu_B$) |
| FM | -301 | 2.28 | -265 | 2.52 |
| AFM1 | -274 | 2.44 | -271 | 2.43 |
| AFM2 | -341 | 2.59 | -345 | 2.59 |
| AFM3 | -348 | 2.59 | -356 | 2.57 |

TABLE II: The results of structural optimization of FeTe in the AFM3 state of the monoclinic (M) and tetragonal (T) phases as well as in the NS state. Structural parameters $a, b, c$ and the distance $h_{Te}$ are given in Å, while $z_{Te}$ is the internal Te structural parameter. The experimental parameters for M and T phase are also shown [7, 15].

| lattice parameters | $a$ | $b$ | $c$ | $z_{Te}$ | $h_{Te}$ |
| --- | --- | --- | --- | --- | --- |
| AFM3 (M) | 3.86 | 3.67 | 6.77 | 0.27 | 1.80 |
| AFM3 (T) | 3.80 | 3.80 | 6.79 | 0.26 | 1.78 |
| NS | 3.81 | 3.81 | 6.50 | 0.25 | 1.62 |
| exp. (M) | 3.83 | 3.78 | 6.26 | 0.28 | 1.75 |
| exp. (T) | 3.82 | 3.82 | 6.29 | 0.28 | 1.76 |

### III. RESULTS

#### A. Magnetism of FeTe under various conditions

The relative values of total energy of spin-polarized (SP) states with respect to the non-spin-polarized (NS) one, $\Delta E = E_{tot}(SP) - E_{tot}(NS)$, obtained for the tetragonal crystal structure with experimental lattice parameters and atomic positions, are collected in table I. The ground state of FeTe is the in-plane AFM3 configuration, while in the consecutive layers the Fe magnetic moments alternate their directions. The antiparallel alignment of the Fe spin moments, repeated along the $c$ axis, does not change energies significantly for most configurations except for the FM in-plane one. However, the latter structure has the smallest stabilization energy, $|\Delta E|$, so it is not important in further calculations. Thus, we consider here only in-plane spin arrangements. The values of magnetic moments of Fe atoms, which are also given in table I, depend on a spin structure and are about 2.3-2.6$\mu_B$/Fe, being in good agreement with the measured values of 2.25-2.54$\mu_B$ in iron-rich Fe$_{1+x}$Te samples [7, 24].

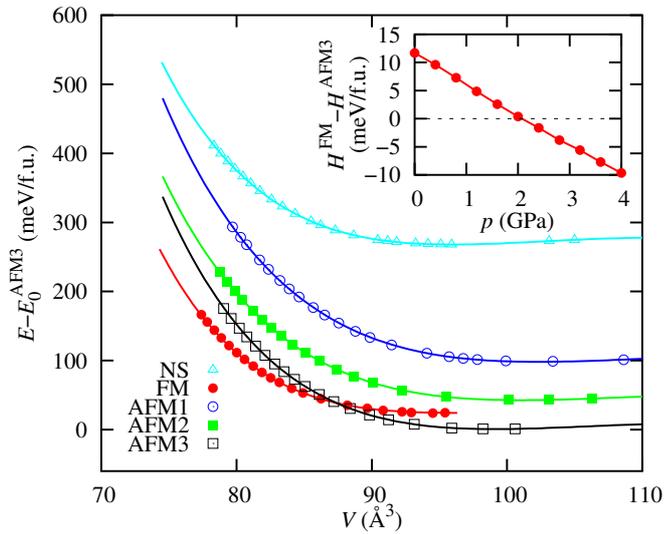

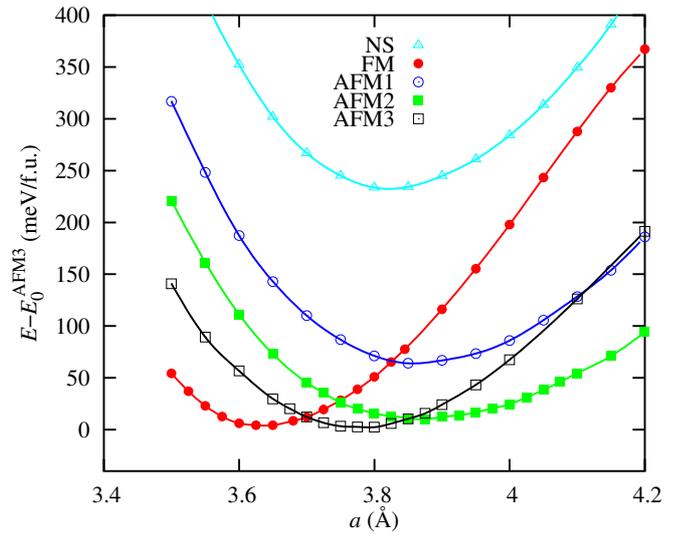

FIG. 2: The volume dependence of the relative total energy per formula unit with respect to the equilibrium AFM3 structure of FeTe in the NS, FM, AFM1, AFM2 and AFM3 states. The inset shows the difference between FM and AFM3 enthalpy in the region of the phase transition.

FIG. 4: The total energy of the tetragonal FeTe in the NS, FM, AFM1, AFM2 and AFM3 states in relation to the AFM3 ground state as a function of the lattice parameter $a$.

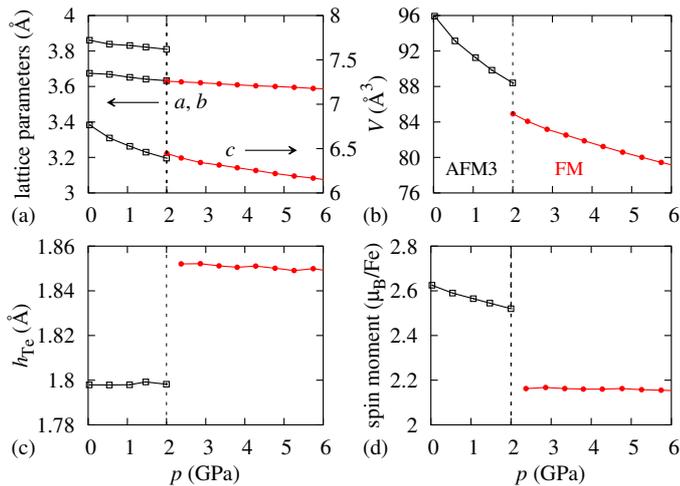

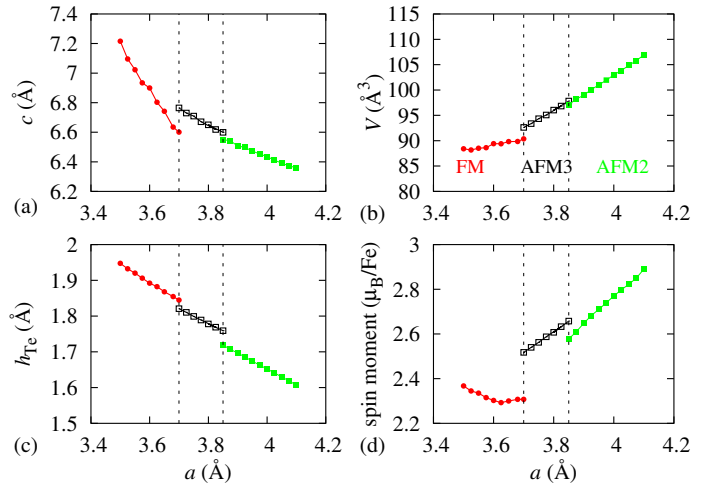

FIG. 3: The structural changes of FeTe under pressure: (a) the lattice parameters $a$, $b$, $c$, (b) the volume of the unit cell, $V$, and (c) the Te distance from the iron plane, $h_{Te}$. (d) The dependence of the Fe magnetic moment on pressure, $p$. The phase transition from the monoclinic AFM3 to tetragonal FM at $p = 2$ GPa is indicated by dashed lines.

FIG. 5: The changes of (a-c) structural parameters of FeTe and (d) spin moment on one Fe atom under biaxial strain as a function of the lattice parameter $a$.

During relaxation under pressure, the AFM2 unit cell distorts to the orthorhombic crystal structure (space group $Cmma$), whereas the AFM3 to the monoclinic one (P2$_1$/m) with the lattice parameters $a \neq b$ and the angle between $a$ and $c$ axes smaller than 90°. The resulting lattice parameters are slightly shorter along spin-parallel alignment for both cases indicating that structural distortions are driven by magnetic interactions [30].

For the ground state (AFM3) of FeTe, the optimized crystal structure parameters at ambient pressure and the corresponding experimental data are collected in table II. Although the lattice parameter $c$ is overestimated within the GGA approximation, the atomic position $z_{Te}$ (and $h_{Te} = z_{Te}c$), which defines the distance between the Te and Fe planes, agrees reasonably well with the experimental values reported in [7, 15]. In turn, the optimization of the NS phase gives a more accurate $c$ value but an underestimated $h_{Te}$ distance.

The volume dependences of the total energy of FeTe possessing different spin configurations are shown in figure 2. The volume of the equilibrium AFM3 state is found to be 96 Å$^3$, *i.e.* a little more than the experimental value of about 92-93 Å$^3$ [15, 33]. However, when the volume contracts, the FM state becomes more stable than the AFM3. Thus, we expect that the magnetic transition takes place under hydrostatic pressure. The critical pressure is determined by condition of equal enthalpies $H = E+pV$ for both phases. The obtained value of 2 GPa (see inset to figure 2) is in good agreement with experimental measurement for Fe$_{1.03}$Te [17].

The spin transition is accompanied by the structural one, since the AFM3 and FM phases occur in the monoclinic and tetragonal lattices, respectively. In addition, the pressure dependences of both lattice parameters and the volume are not continuous functions at $p =2$ GPa (figure 3). Hence, such a transition might be classified as of the first order. It is worth noting that the jump in $V(p)$ is not obtained within non-magnetic calculations, so it cannot be observed experimentally at room temperature [33, 34]. Similarly, the discontinuity is found neither in FM nor AFM single phases. It occurs as a consequence of the structural transition caused by shortening spin-parallel Fe-Fe bonds. Precisely, it is a result of contraction of the lattice parameter $a$, which corresponds to the antiparallel Fe spin arrangement in the AFM3 state, but to the parallel arrangement in the FM state. The $c$ parameter decreases faster than $a$ and $b$ with increasing pressure, as in the experiment [33, 34]. In turn, the distance between the Te and Fe-plane, $h_{Te}$, is nearly independent of pressure within one magnetic phase, but abruptly raises from about 1.80 Å to 1.86 Å going from the AFM3 to FM phase. It significantly contracts the interlayer Te-Te distance, *i.e.* the distance in the FM phase at $p =2$ GPa is about 10% shorter than the one in the AFM3 phase at $p =0$. For comparison, the contraction of Fe-Fe and Fe-Te bonds is below 4 and 0.3%, respectively. The obtained effects require a verification by low-temperature measurements.

The next difference between the two magnetic phases is visible for the spin moment on single Fe atom (figure 3d). In the AFM3 state, the moment decreases monotonically with increasing pressure, while in the FM state, the moment on the Fe atom shows the drastic drop and then is nearly constant. The result is inconsistent with the experimental data, where the Fe magnetic moment increases and reaches a value of 3$\mu_B$ [17].

Move on to discuss the effect of $ab$-plane stress on magnetism of FeTe. Figure 4 depicts the dependence of total energy of the system having different spin configurations on the lattice parameter $a$. The ground state of the AFM3 phase fixed to the tetragonal crystal structure is realized for $a = 3.80$ Å. Some characteristics of this state are also given in table I. For compression of the lattice parameter of 3% ($a$ =3.70 Å), the transition from the AFM3 to FM states takes place, similar to that under hydrostatic pressure. For tension of about 2% ($a$ =3.87 Å), the AFM2 phase emerges as a ground state.

Figure 5 shows the changes of structural parameters for the considered system. The lattice parameter $c$ decreases with increasing $a$, but the changes are slower under tension than under compression. Thus, the AFM3-AFM2 switch is accompanied by a softening of the first-order structural phase transition [22] . The $c$ parameter, though, is not reproduced properly either at equilibrium (table II) or under pressure conditions. For superconducting thin films, the lattice parameter $c$ is less than 0.5% shorter than in the bulk FeTe.

The $h_{Te}$ distance also decreases with $a$. The result of compressive strain is analogous to the hydrostatic pressure, where the transition from the AFM3 to FM phases happens for $h_{Te}$ high enough. Similarly, transition from the AFM3 to AFM2 phases under tensile strain is accompanied by shortening the $h_{Te}$ distance. The changes of the Fe magnetic moment resemble the ones under hydrostatic pressure, too. In the FM state, the moment on the Fe atom is the lowest and nearly independent of continued compression.

The in-plane strain yields the increase of the distance between the Te and Fe planes with decreasing $a$-axis (figure 5(c)) and causes two magnetic phase transitions: the AFM3-FM transition for $h_{Te} > 1.84$ Å and AFM3-AFM2 transition for $h_{Te} < 1.72$ Å. In contrast, the AFM2 state is never the most stable under hydrostatic pressure (figure 3). The lattice parameter $a$ decreases weakly on $p$ and simultaneously the tellurium position is nearly independent of $p$ in both AFM3 ($h \approx 1.80$Å) and FM phase ($h \approx 1.87$Å). On the one hand, it points to the correlation between the parameters $a$ and $h_{Te}$ in FeTe. On the other hand, the jump of $h_{Te}$ under pressure is driven by the structural transition occurring at 2 GPa rather than by the in-plane strain.

Moreover, the superconducting Fe(Te,Se) alloys also favour the AFM2 ground state as follows from previous DFT calculations [30, 31]. In this case, the Se substitution for Te leads to both $a$-axis and $h_{Te}$ distance compression [24–26]. Thus, the influence of the $a$-lattice parameter on the $h_{Te}$ distance and the magnetic ground state is rather complex and not universal for iron chalcogenides. Nevertheless, the chalcogen position $h_{Te}$ seems to be the key structural parameter that drives the magnetic phase transitions in the Fe(Te,Se) systems.

Therefore, we check if the Te position changes alone can reproduce a similar phase diagram. For convenience, the tetragonal crystal structure with experimental lattice parameters $a$ and $c$ is assumed [15]. Figure 6 shows the dependence of total energy of FM and all AFM states on $z_{Te}$. A similar analysis was done in the previous work [31] to show the AFM2-AFM3 switch going from FeSe to FeTe. This transition, obtained for $z_{Te} \cong 0.27$, is also present in figure 6 and reproduces our result for FeTe under tensile strain. Interestingly, for a $z_{Te}$ value lower than 0.23, only the AFM2 phase has small stabilization energy, $|\Delta E| = 20$ meV, and for $z_{Te} < 0.22$ the system becomes non-magnetic (see the inset to figure 6). On the other hand, for a high enough value of $z_{Te}$, the switch from the AFM2 to FM phases takes place, according to our spin transitions under pressure or compressive strain. Thus, the complete magnetic phase diagram for tetragonal FeTe with increasing $z_{Te}$ is as follows: NS, AFM2, AFM3 and FM.

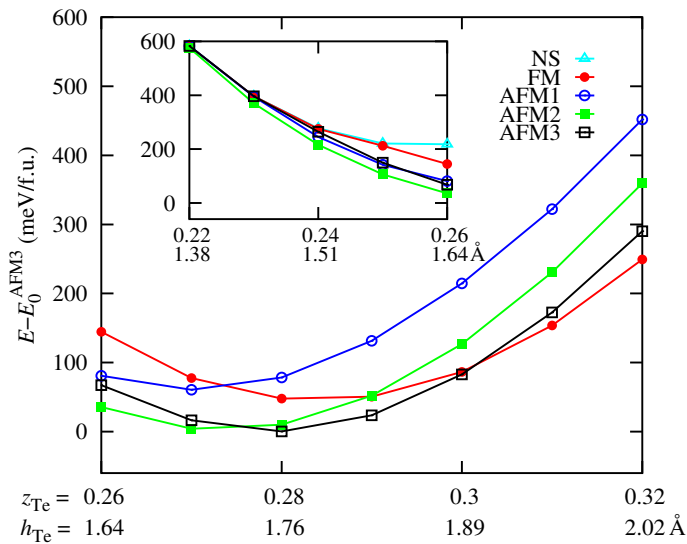

FIG. 6: (a) The relative total energy with respect to the AFM3 ground state ($z_{Te}$ =0.279) state vs. $z_{Te}$ calculated for the NS, FM, AFM1, AFM2 and AFM3 phases for experimental values of parameters $a$ and $c$ [15].

### B. Pressure effect on electronic properties of FeTe

In this section, the $ab$-plane-strained FeTe systems are considered because of three possible magnetic phases. The differences between the AFM3 and FM states under hydrostatic pressure and even between all states in the model presented in figure 6 are qualitatively the same.

To examine the electronic structure of FeTe, bands and the densities of states (DOS) of non-magnetic phases for a few values of the lattice parameter $a$ are shown in figure 7. The $c$ and $z_{Te}$ parameters are taken from optimization of magnetic phases, stable for given $a$, because the AFM3 unstrained system better reproduces experimental data

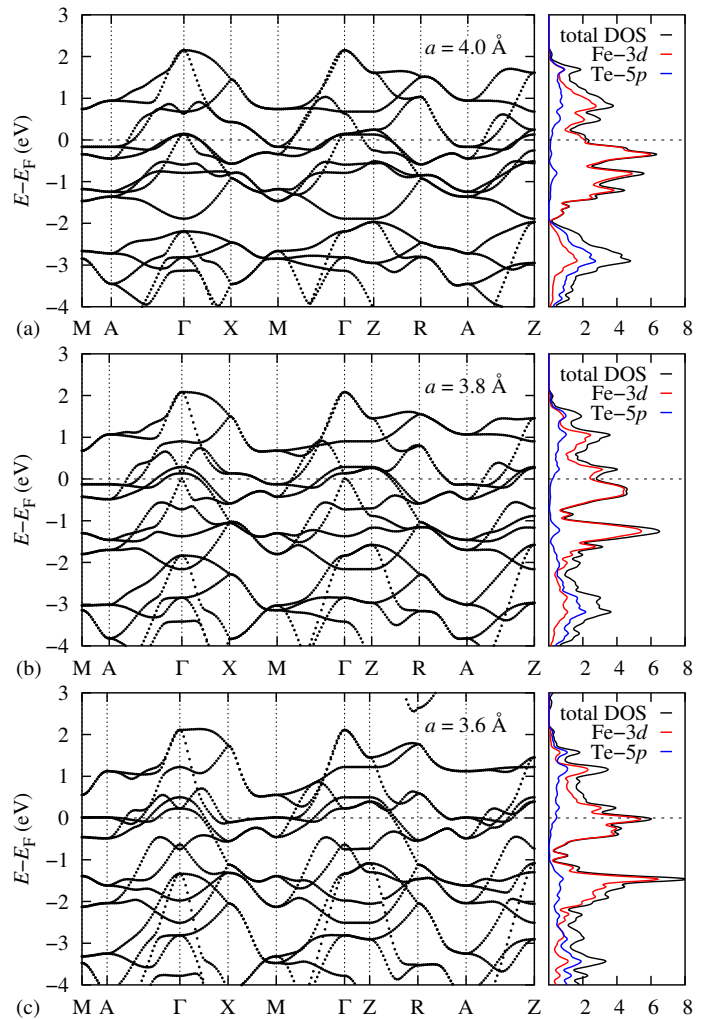

FIG. 7: Non-spin polarized bands and DOS of FeTe under $ab$-plane strain: (a) $a$ = 4.0 Å, (b) $a$ =3.8 Å, (c) $a$ =3.6 Å. The structural parameters $c$ and $z_{Te}$ are taken from relaxation of the spin polarized state, i.e. AFM2, AFM3 and FM, respectively.

for the $h_{Te}$ distance than the NS system (see table II).

In general, the results for unstrained FeTe agree with the previous theoretical works [5, 31], some bands are somewhat redistributed as we used the fully optimized lattice parameters in the AFM3 phase. The states near the Fermi energy consist of the Fe-3$d$ orbitals slightly hybridized with the chalcogenide 5$p$ oritals. For FeTe under tensile stress (figure 7a), there is a pseudogap, which substantially reduces the DOS at the Fermi level, $E_F$, similarly to the case of superconducting Fe(Se,Te) systems [5, 35, 36]. The size of the pseudogap is significantly smaller for unstrained FeTe and FeTe under biaxial strain (figure 7b, c). Additionally, the Fermi energy shifts closer to the peak in the DOS. As a result, the value of DOS at $E_F$ increases quickly with decreasing the $a$ parameter. It equals 2.2, 2.8 and 5.7 states/eV per both spins for $a = 4.0$, 3.8 and 3.6 Å, respectively. It provides conditions for an appearance of the Stoner instability and suggests that an itinerant ferromagnetism is realized in FeTe.

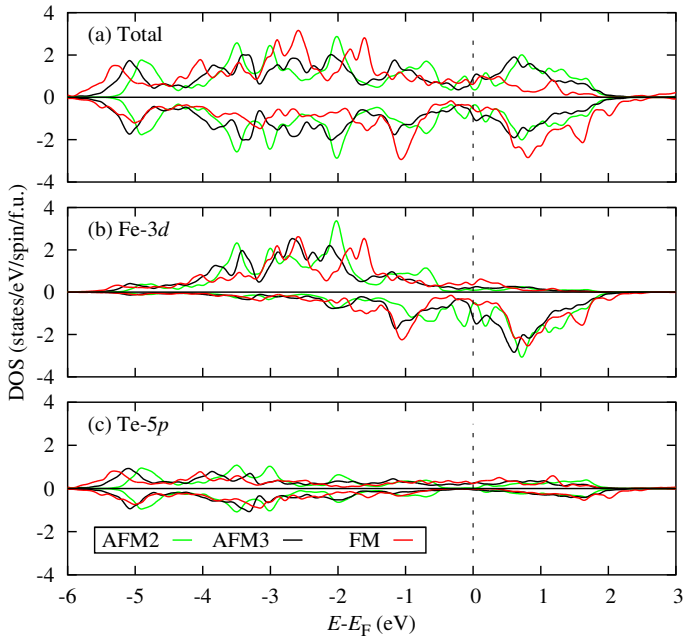

FIG. 8: Spin polarized (a) total and partial DOS of (b) Fe-3$d$, (c) Te-5$p$ orbitals under tensile stress ($a = 4.0$ Å, AFM2 phase), unstrained ($a = 3.8$ Å, AFM3 phase) and under compressive stress ($a = 3.6$ Å, FM phase).

Although the changes are substantial mainly for the Fe-3$d$ states, the partial DOS of the Te-5$p$ orbitals at $E_F$ also increases with decreasing $a$. Since the interlayer Te-Te distance, $d_{Te-Te}$, is reduced under compression, the more itinerant Te electrons result in the higher charge density mainly between the Fe-Te layers.

The DOS's for magnetic phases of FeTe are shown in figure 8. For the AFM2 and AFM3 phases, the majority spin states of the Fe-3$d$ orbitals are nearly completely occupied, while the minority spin states are only partially filled (figure 8b). They are almost uniformly distributed into the five 3$d$ orbitals (not shown) indicating strong hybridization of the Fe orbitals and possible Hund's rule coupling [30]. As a result, the magnetic moments on the Fe atom of 2.5-2.6$\mu_B$ in both AFM states are higher than that in the FM state of 2.2$\mu_B$ (see also figure 5). In turn, the DOS projected onto the Te-5$p$ orbitals (figure 8c) indicates that the induced magnetic moment on the Te atom is present in the system in the FM state. In the AFM states, the calculated magnetic moment on the Te atom is close to zero, whereas in the FM state it amounts to about -0.1$\mu_B$/Te. The Te-5$p$ states, contrary to the Fe-3$d$ ones, are spreaded in a wider range of energy, which can point to a stronger Te-Te interlayer hybridization.

## IV. DISCUSSION AND CONCLUSIONS

In the considered FeTe system, two possible spin phase transitions have been obtained under various stress condition. Our result of the AFM3-FM switch under hydrostatic pressure of 2 GPa was recently confirmed [17]. A similar transition is found to appear also under in-plane compressive strain. Under tensile strain the single stripe AFM2 state becomes more stable than the AFM3 state, while the superconductivity was detected in epitaxial films [22]. It suggests that the superconducting state is positively linked to the single-stripe antiferromagnetic fluctuations.

Indeed, the correlation between antiferromagnetic fluctuation and superconductivity was found to exist in Fe(Te,Se) samples. The parent compound FeTe shows an antiferromagnetic order with the $(\pi, 0)$ propagation vector, whereas the superconducting state obtained by the Se substitution for Te atoms displays a suppression of the spin resonance at the $(\pi, 0)$ wave vector and an enhancement at the $(\pi, \pi)$ wave vector [37–40].

In the real samples, AFM fluctuations are also dependent on stoichiometry related to an excess Fe at interstitial sites of the Te layers. The out-of plane Fe ions are strongly magnetic and their local moments interact with the in-plane Fe magnetism [6]. They cause turning from commensurate into incommensurate spin fluctuations in the iron-rich samples [8–10, 41]. As an effect, the superconductivity does not occur even in the Se substituted samples. The results of presented calculations are based on the assumption that no-extra Fe site are occupied in the FeTe structure. However, the AFM3-FM switch under pressure [17] is primarily the effect of in-plane changes of Fe properties, as it is well reproduced in the pure FeTe system. Similarly, we expect that the superconductivity in thin film [22] occurs as an effect of in-plane AFM2 fluctuations and is possible only in nearly stoichiometric FeTe.

Both hydrostatic and biaxial pressure have a substantial effect on the crystal structure, predominantly on the distance between the Te and Fe planes. It affects electronic and magnetic properties of the system, which we checked also for the model of the fixed tetragonal unit cell of FeTe and $z_{Te}$ as a variable parameter.

For thin films with low value of $h_{Te}$, the DOS at the Fermi level decreases because of the pseudogap, typical of semimetals. The Fe magnetic moment increases with decreasing $h_{Te}$ and is sensitive to spin ordering. It suggests that an itinerant magnetism is realized in the system [5]. The AFM2 state and $(\pi, \pi)$ spin fluctuations become more preferable, which coincides with the superconducting state.

For FeTe at equilibrium, double stripe AFM with the $(\pi, 0)$ ordering vector is realized [7, 8]. The range of $h_{Te}$, where the AFM3 phase is stable, is rather narrow and the results show that the system is on the verge of ferromagnetic instability. For a little higher values of $h_{Te}$, the DOS at the Fermi level rapidly raises and the system turns into the FM state. Nonetheless, the authors of [17] suggest the localized nature of the magnetism in FeTe samples.

Indeed, the Curie-Weiss behaviour of the spin susceptibility is observed above the Neel temperature at ambient pressure [42, 43], which is an experimental evidence of the strong local magnetism in FeTe. Therefore, the series of theoretical works imply the Heisenberg-type models to explain the magnetism of FeTe [30–32]. We also examined the superexchange interactions to obtain both AFM3-AFM2 and AFM3-FM transitions. The dependence of the total energy can be easily mapped onto suitable short-range exchange interactions, but the model seems to be unreliable (see remark in [44]), which was also discussed widely in the reference [45].

Moreover, FeTe is metallic and its itinerant electrons may coexist with the Fe local moments. Our calculation results indicate that the higher DOS at the Fermi level observed for FeTe under pressure in the paramagnetic state is an important feature. Thus, the more effective way to describe iron chalcogenides seems to be models with both itinerant electrons and on-site correlations driven by Hund's rule coupling [45]. The intra atomic exchanges are consistent with the raise of our DOS at $E_F$ with the $h_{Te}$ distance on the one site, and are of a local type on the other site.

Irrespectively of the right picture of magnetism in FeTe, the transition from an antiferromagnetic to ferromagnetic phase, shown in Figs. 2 and 6, explains the absence of a desirable superconducting phase under pressure. The single-stripe AFM arrangement is more stable than double stripe AFM order under negative biaxial pressure realized in epitaxial films (Figs. 4, 6).

There exist optimal distances between Te and Fe planes in the system, which makes the superconducting state possible [13, 14, 31]. The hydrostatic pressure increases chalcogen position, so it is not strange that pressure has a positive effect on critical temperature for FeSe with smaller $z_{Se}$ =1.45 Å, but it does not induce superconductivity in the case of FeTe with $z_{Te}$ =1.76 Å. Here, the more desirable is smaller $z_{Te}$, which can be achieved by partial Se- substitution for Te [2, 3] as well as by *ab*-plane tensile stress in thin film on the appropriate substrate [22].

In summary, the magnetic phase transitions from AFM3 to FM arrangement and from the AFM3 to AFM2 arrangement are obtained from DFT calculations. The first one is achieved under hydrostatic pressure, consistent with the experiment [17], and under *ab*-plane compressive strain. The second one emerges under *ab*-plane tensile strain, whereas the superconducting state was detected in the experiment [22]. The result is in agreement with the scenario that spin fluctuations promote superconductivity in Fe-based chalcogenides and points out that the Te position in the crystal plays a crucial role in both magnetism and superconductivity.

### Acknowledgments

The authors acknowledge Prof. Piotr Bogusławski for valuable discussions. This work has been supported by the EC through the FunDMS Advanced Grant of the European Research Council (FP7 Ideas) as well as by the National Science Centre in Poland under grants No. N N202 239540, No. 2011/01/B/ST3/02374, and No. 2012/05/B/ST3/03095. Calculations were partially performed on ICM supercomputers of Warsaw University (Grant No. G46-13) and in Wrocław Centre for Networking and Supercomputing (Project No. 158).